\documentclass[sigconf]{acmart}

\usepackage[utf8]{inputenc} 
\usepackage[T1]{fontenc}    
\usepackage{hyperref}       
\usepackage{url}            
\usepackage{amsfonts}       
\usepackage{nicefrac}       
\usepackage{microtype}      
\usepackage{xcolor}         
\usepackage{colortbl}

\usepackage{algpseudocode}

\usepackage{framed}
\usepackage{makecell}
\usepackage{multirow}
\usepackage{amsthm}
\theoremstyle{plain}

\theoremstyle{definition}

\theoremstyle{remark}

\usepackage{tabularx}

\newcommand{\delete}[1]{} 

\definecolor{cblue}{RGB}{218,232,252}
\definecolor{cgreen}{RGB}{213,232,212}
\definecolor{cyellow}{RGB}{255,242,204}
\definecolor{cred}{RGB}{248,206,204}
\definecolor{corange}{RGB}{255,230,204}
\definecolor{cgray}{gray}{.9}

\AtBeginDocument{%
  \providecommand\BibTeX{{%
    \normalfont B\kern-0.5em{\scshape i\kern-0.25em b}\kern-0.8em\TeX}}}

\setcopyright{acmcopyright}
\copyrightyear{2018}
\acmYear{2018}
\acmDOI{XXXXXXX.XXXXXXX}

\acmConference[Conference acronym 'XX]{Make sure to enter the correct
  conference title from your rights confirmation email}{June 03--05,
  2018}{Woodstock, NY}
\acmPrice{15.00}
\acmISBN{978-1-4503-XXXX-X/18/06}

\begin{document}

\title{Towards Dynamic Dense Retrieval with Routing Strategy}

\author{Zhan Su}
\affiliation{%
  \institution{Université de Montréal}
  \city{Montréal}
  \state{Québec}
  \country{Canada}
}
\email{zhan.su@umontreal.ca}

\author{Fengran Mo}
\affiliation{%
  \institution{Université de Montréal}
  \city{Montréal}
  \state{Québec}
  \country{Canada}
}
\email{fengran.mo@umontreal.ca}

\author{Jinghan Zhang}
\orcid{0009-0001-0999-270X}
\affiliation{%
  \institution{Clemson University}
  \city{Clemson}
  \state{South Carolina}
  \country{USA}
}
\email{jinghaz@clemson.edu}

\author{Yuchen Hui}
\orcid{0000-0002-9659-3714}
\affiliation{%
  \institution{Université de Montréal}
  \city{Montréal}
  \state{Québec}
  \country{Canada}
}
\email{yuchen.hui@umontreal.ca}

\author{Jia Ao Sun}
\orcid{}
\affiliation{%
  \institution{Université de Montréal}
  \city{Montréal}
  \state{Québec}
  \country{Canada}
}
\email{jia.ao.sun@umontreal.ca}

\author{Bingbing Wen}
\affiliation{
\institution{University of Washington}
\country{America}
}
\email{bingbw@uw.edu}

\author{Jian-Yun Nie}
\affiliation{%
  \institution{Université de Montréal}
  \city{Montréal}
  \state{Québec}
  \country{Canada}
}
\email{nie@iro.umontreal.ca}


\begin{abstract}
    The \textit{de facto} paradigm for applying dense retrieval (DR) to new tasks involves fine-tuning a pre-trained model for a specific task. However, this paradigm has two significant limitations: (1) It is difficult adapt the DR to a new domain if the training dataset is limited. 
    (2) Old DR models are simply replaced by newer models that are trained from scratch when the former are no longer up to date. Especially for scenarios where the model needs to be updated frequently, this paradigm is prohibitively expensive. To address these challenges, we propose a novel dense retrieval approach, termed \textit{dynamic dense retrieval} (DDR). DDR uses \textit{prefix tuning} as a \textit{module} specialized for a specific domain. These modules can then be compositional combined with a dynamic routing strategy, enabling highly flexible domain adaptation in the retrieval part. Extensive evaluation on six zero-shot downstream tasks demonstrates that this approach can surpass DR while utilizing only 2\% of the training parameters, paving the way to achieve more flexible dense retrieval in IR. We see it as a promising future direction for applying dense retrieval to various tasks. 
\end{abstract}





\maketitle

\section{Introduction}

Dense retrieval approaches have shown remarkable capabilities, representing a notable advancement from classical retrieval methods \citep{karpukhin2020dense,lin2020distilling,qu2020rocketqa,zhao2024dense,mo2024history,xiong2020approximate} to neural retrieval \citep{gao2021condenser}. By embedding queries and documents into a latent vector space using dual-encoders, dense retrieval methods have demonstrated remarkable success in tasks such as web search \citep{kim2022applications}, question answering \citep{karpukhin2020dense}, and recommendation systems \citep{huang2024comprehensive}. In web search, DR has enhanced the ability of search engines to deliver more relevant results by understanding the intent behind user queries \citep{zhu2021retrieving}. In question answering, DR bridges the semantic gap between queries and candidate answers by embedding both 
into a shared semantic space, allowing for more accurate retrieval of contextually relevant information \citep{dpr,roy2022question,huang2024comprehensive}.

Despite their success, current DR approaches still face two challenges: \textbf{(1)} Once DR models are trained and fixed, it is difficult to adapt them to a new domain when training data is limited. A common approach is to build domain-specific datasets and fine-tune the retriever on them \citep{zhao2024dense}. However, collecting and annotating such data is often difficult and expensive.  \textbf{(2)} In scenarios where dense retrieval models need frequent updates, older models are often replaced by new models trained from scratch. This wastes significant training resources. These two challenges lead to a natural question: \textit{Can we develop a dynamic dense retrieval approach that can adapt to new domains quickly while fully leveraging previously trained models?}

To answer this question, we propose dynamic dense retrieval (DDR) with a fine-grained routing function.  In this approach, we use prefix tuning \citep{li2021prefix} to adapt dense retrieval models to new domains efficiently. Prefix tuning prepends $l$ trainable prefix vectors to the pre-trained model. During training, the parameters of the backbone remain frozen, and only the parameters associated with the prefix vectors are updated. We treat each \textit{prefix vector} as an \textit{module} specialized in a specific domain (\S\ref{sec:modularization_prompt}). Then a routing function can dynamically select which module should be activated in the retrieval part (\S\ref{sec:routing}). This approach enables rapid adaptation to new tasks, domains, or emerging topics without requiring training from scratch. In addition, DDR can leverage previously trained modules for new tasks, rather than simply discarding old dense retrieval models, thereby significantly reducing training cost and improving knowledge reuse.

We evaluate DDR on downstream retrieval tasks that require domain composition. The experimental results demonstrate that DDR can surpass traditional dense retrieval models while utilizing only 2\% of training parameters.

To sum up, our contributions are as follows: 

\begin{itemize}
    \item We formally introduce a new dense retrieval approach: \textit{Dynamic dense retrieval}, converting dense models into modular models in the field of information retrieval. DDR addresses the limitations of traditional dense retrieval systems by offering flexibility, scalability, and efficiency. 
   
    \item We propose fine-grained routing strategies, which dynamically activate the prefix vectors during retrieval, resulting in a dynamic dense retrieval system capable of adapting to new tasks.

    \item The experimental results indicate that DDR can surpass the traditional DR while only using 2\% training parameters. Meanwhile, the DDR can leverage the previously trained modules for new tasks and avoid discarding old dense retrieval models, thereby reducing training cost and improving knowledge reuse. The code is released: \footnote{\url{https://anonymous.4open.science/r/REMOP-sigir2026-short/README.md}}
    
\end{itemize}


\begin{figure*}[htp]
    \centering
    \includegraphics[width=0.9\textwidth]{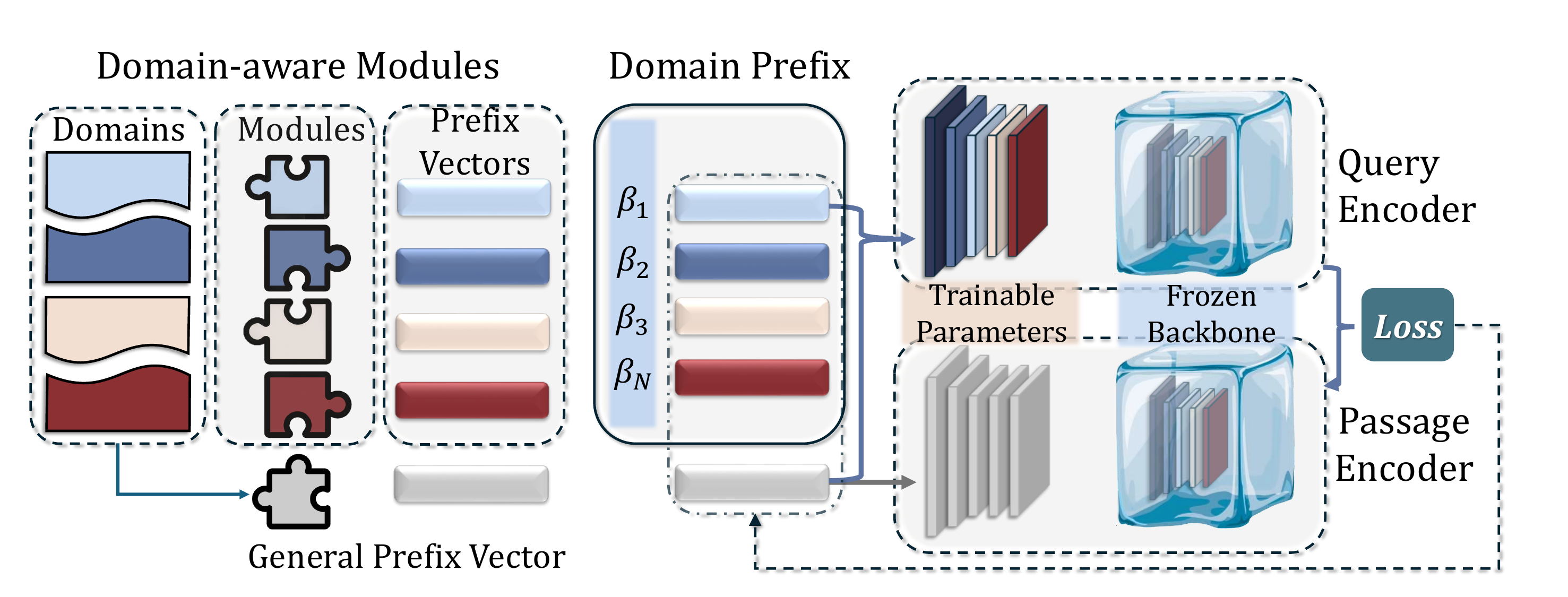}
    \caption{Overview of DDR framework, built upon a dual-encoder architecture. The general prefix vector learns the general knowledge across tasks, while the domain prefix focuses on the specificity of each task domain. The frozen backbone incorporates both the general prefix and aspect prefix for query encoding, whereas the passage encoder only utilizes the general prefix.}
    \label{fig:framework of REMOP}
\end{figure*}
\section{Related Work}
\label{section: related works}

Dense Retrieval (DR) models convert queries and documents into dense embeddings through which documents are selected \citep{peng2025soft,zhan2021optimizing,kong2022multi,zhao2024dense,lin2023train,ma2024fine,izacard2021unsupervised,chen2022salient}. The dense passage retrieval (DPR) model employs a two-tower architecture based on BERT to encode the query and document separately \citep{dpr}. Colbert \cite{khattab2020colbert} utilizes one BERT encoder by concatenating the query and the document as the input, and outputs a similarity score. 
ANCE \cite{xiong2020approximate} is a bi-encoder trained on (query, positive document, negative document) tuples where the negative document is retrieved from an ANN built on the checkpoint of the last step. Contriever \cite{izacard2022unsupervised} trains a bi-encoder model through contrastive learning. CoCondenser \cite{gao2021unsupervised} introduces a novel pre-training architecture, which learns to condense information into a dense vector through LM pre-training. Ma, Xueguang, et al. \cite{ma2024fine} finetune an open-source LLaMA-2 model as a dense retriever. However, both dense retrieval models have been trained independently from scratch. For a new task or domain, one needs to retrain a new model with domain-specific data. There are no principled methods capable of combining them in a flexible way to take advantage of different models' capabilities. Our contribution precisely lies in 
a new paradigm to enhance the flexibility of dense retrieval by combining the available models.

\section{Preliminary}

\subsection{Prefix Tuning}
\label{sec:prompt_tuning}
Multi-head attention performs the attention function in $N_h$ heads, where each head is separately parametrized by ($w_q^{(i)},w_k^{(i)},w_v^{(i)}\in\mathbb{R}^{d\times d_h}$) \cite{he2021towards}. Given a sequence of $m$ vectors $C\in\mathbb{R}^{m\times d}$ over which we perform attention, and a query vector $x\in\mathbb{R}^d$,
Prefix tuning \cite{li2021prefix} prepends $l$ tunable prefix vectors to the keys and values of the multi-head attention at each layer. Two sets of prefix vectors $P_k$ and $P_v$ are concatenated with the original key $W_k$ and value $W_v$. Then the computation of head becomes:
\begin{equation}
    \mathrm{head}_{(i)}=\mathrm{Attention}(xW_q^{(i)},[ P_k^{(i)}; CW_k^{(i)}],[P^{(i)}_v;CW_v^{(i)}])
\end{equation}
where $[;]$ is the concat operation, $P_k$ and $P_v$ are the split into $N$ heads and $P_k^{(i)}$ and $P_v^{(i)}$ denote the $i$-th head vector.

\section{DDR: Dynamic Dense Retrieval with Routing Strategy}
\label{section: retrieval using deep modular prompts}

In this section, we present the details of DDR. Specifically, we first introduce dynamic dense retrieval training based on prefix vector tuning and routing strategy. $\S \ref{sec:modularization_prompt}$. Then we compare the various routing strategy in ($\S$ \ref{sec:routing}).



\subsection{Dynamic Dense Retrieval Training}
\label{sec:modularization_prompt}
As shown in Figure \ref{fig:framework of REMOP}. DDR adopts a dual-encoder architecture consisting of a query encoder $f_q$ and a passage encoder $f_p$, which project queries and passages into a shared embedding space. 
We introduce two types of modules: a general module, which captures domain-agnostic retrieval knowledge, and a domain module, which encodes domain-specific information. In practice, dense retrievers pre-compute passage embeddings and build an approximate nearest neighbor (ANN) index for efficient retrieval \citep{peng2025soft}. To minimize storage overhead and enable reuse of indexed passages, we incorporate the general module into both the query and passage encoders, while attaching the domain module only to the query encoder. This design allows a single set of passage embeddings to be shared across domains, making DDR efficient to deploy and easy to adapt to new tasks. 

Let the encoder consist of $L$ transformer layers, each with $H$ attention heads, and the head dimension $d$. We define the general prefix vector $\mathcal{P}^g=\{P_k^g,P_v^g\}$ and domain prefix vector 
$\mathcal{P}^d=\{P_k^d,P_v^d\}$. Then the prefix vector of the passage encoder $f_p$ in each transformer layer is :

\begin{equation}
    \mathcal{P}^{(passage)} = \mathcal{P}^g
\end{equation}

For the query encoder, DDR uses both the general prefix vector and the domain prefix vector. To allow DDR to dynamically choose the required domain during retrieval, we introduce a selection mechanism, which is implemented through a routing function $r(\cdot)$. Given the input $x_l$ at layer $l$, the router produce logits $z_l=W_rx_l$ with routing parameters $\Delta_r=W_r\in\mathbb{R}^{N\times d}$, then the routing distribution is computed by:

\begin{equation}
    \beta_i =r(x_l)=\text{softmax}(z_l)
\end{equation}

Then the prefix vector of the query encoder $f_q$ in each transformer layer is:
\begin{equation}
\label{eq:query_prefix}
    \mathcal{P}^{(query)} = \mathcal{P}^g+\sum_i^N\beta_i\cdot \mathcal{P}^d
\end{equation}
where  $P_v^g\in \mathbb{R}^{H\times L_g\times d}$ and $P_k^d,P_v^d\in\mathbb{R}^{H\times L_d\times d}$ and $L_g$ and $L_d$ denote the prefix lengths; $N$ indicates the total number of domain. The attention output for the query encoder and passage encoder is computed as:
\begin{equation}
\begin{split}
\mathrm{head}_{(q)}&=\mathrm{Attention}(xW_{q},[\mathcal{P}^{(query)}_k, CW_k],[\mathcal{P}^{(query)}_v, CW_v])\\
\mathrm{head}_{(p)}&=\mathrm{Attention}(xW_{p},[\mathcal{P}^{(passage)}_k, CW_k],[\mathcal{P}^{(passage)}_v, CW_v])
\end{split}
\end{equation}
Then we can compute the final query presentation and passage representation as follows:

\begin{equation}
\begin{split}
      z_q&=f_q(q;\mathcal{P}^g,\mathcal{P}^d) \\
      z_p&=f_p(p;\mathcal{P}^g)
\end{split}
\end{equation}

Given a query-passage pair, the relevance score is computed by the inner product:

\begin{equation}
    s(q,p)=z_q^Tz_p
\end{equation}

We train the DDR using a contrastive objective:

\begin{equation}
\label{eq:training_eq}
    \mathcal{L}_{\text{DDR}} = \min_{\mathcal{P}^g,\mathcal{P}^d,W_r}-\frac{1}{n} \sum_{i=1}^n \log \frac{\exp(s(q_i, p_i^+))}{\sum_{j=1}^K \exp(s(q_i, p_j))}
\end{equation}
where $\mathcal{P}^g,\mathcal{P}^d$ are the training parameters of general prefix vectors and domain prefix vectors, $W_r$ is the training parameters of the routing function, and $p_i^+$ denotes a positive sample.


\subsection{Routing Strategy Comparison}
\label{sec:routing}

The routing function in Eq.\ref{eq:query_prefix} is a soft routing distribution, in which all modules will be activated. In this subsection, we discuss two sparse routing, which only activate a subset of modules. 

\subsubsection{Top-$k$ routing (DDR-topk)}
Inspired by the top-$k$ routing function from the mixture of experts \cite{jiang2024mixtral}, we can activate only the top prefix vectors based on the routing scores. Then the prefix vector in the query encoder $f_q$ is:
\begin{equation}
\label{eq:query_prefix_topk}
    \mathcal{P}^{(query)} = \mathcal{P}^g+\sum_i^K\mathbb{I}[i\in \mathbf{topk}(\beta)]\cdot\beta_i\cdot\mathcal{P}^d
\end{equation}

\subsubsection{Routing with prior information (DDR-prior)}

Instead of top-$k$ routing, we can activate a subset of prefix vectors using prior information. As shown in Table~\ref{tab:task_extraction}, the training instructions can be decomposed into multiple domains, and we can use these prelabelled \textit{prior} signals. Based on the labeled instruction, we select the corresponding prefix vectors. The resulting prefix vector used in the query encoder $f_q$ is:

\begin{equation}
  \label{eq:query_prefix_prior}
    \mathcal{P}^{(query)} = \mathcal{P}^g+\sum_i\mathbb{I}[i\in \mathcal{M}(\beta)]\cdot\beta_i\cdot\mathcal{P}^d  
\end{equation}
where $\mathcal{M}(\cdot)$ is a instruction-domain mapping function. In this paper, we use the pre-labeled instruction datasets used in the BERRI \cite{asai2022task}. Each instruction is mapped into several domains. 

\begin{table}[h]
\centering
\small
\caption{Examples of instruction decomposition for retrieval tasks in the BERRI \cite{asai2022task}. Each task has diverse domains, which are denoted by experts.}
\begin{tabular}{p{2cm}p{5cm}}
\hline
\textbf{Task} & \textbf{Instruction}                      \\ \hline
NF Corpus        & Retrieve \colorbox{cblue}{scientific paper} paragraph to \colorbox{cgreen}{answer this question.}              \\
SciFact          & Retrieve a \colorbox{cblue}{scientific paper} sentence to \colorbox{cyellow}{verify if the following claim is true.} \\
\hline
\textbf{domains}      & \colorbox{cblue}{Science}, \colorbox{cgreen}{QA}, \colorbox{cyellow}{Fact-Checking}, \colorbox{cred}{Wikipedia}                                         \\ \hline
\end{tabular}
\label{tab:task_extraction}
\end{table}

\begin{table*}[h]
\centering
\vspace{-15pt}
\caption{
Zero-shot retrieval results measured by NDCG@10. \# Params denote the trainable parameters. The highest results among dense retrievers are bolded, and the second-best results are \underline{underlined}. We report the results of baselines from their original papers. The labeled domains for each task are listed in colors: \colorbox{cblue}{Science}, \colorbox{cgreen}{QA}, \colorbox{cyellow}{Fact-Checking}, \colorbox{cred}{Wikipedia}, \colorbox{corange}{Summarization}.
}
\begin{tabular}{llccccccc}
\toprule 
\textbf{Method} &  \textbf{PLM} (\# Params) &  \textbf{TREC} & \textbf{NFC} & \textbf{SCD} & \textbf{SCF}  & \textbf{CLI} & \textbf{DBP} & \textbf{Avg.}  \\ 
 &   &  \colorbox{cblue}{SCI}\colorbox{cgreen}{QA} & \colorbox{cblue}{SCI}\colorbox{cgreen}{QA}  & \colorbox{cblue}{SCI}\colorbox{corange}{SUM}  & \colorbox{cblue}{SCI}\colorbox{cyellow}{FC}  & \colorbox{cyellow}{FC} & \colorbox{cred}{WIKI} & \\
\midrule
\textbf{DeepCT}             & BERT-base (110M) & 40.6 & 28.3 & 12.4 & \underline{63.0} &  6.6 & 17.7 & 28.1\\  
\textbf{SPARTA}             & DistilBERT (66M) & 53.8 & 30.1 & 12.6 & 58.2 &  8.2 & 31.4 & 32.8\\  
\hline
\textbf{Contriever}         & BERT-base (110M) & 27.4 & \textbf{31.7} & \underline{14.9} & \textbf{64.9} & 15.5& 29.2 & 30.6\\

\textbf{ANCE}          & RoBERTa-base (110M)   & \textbf{65.4} & 23.7 & 12.2 & 50.7 & 19.8& 28.1 & 33.3\\
\textbf{DR} & coCondenser (110M) & 63.6 & 28.8 & 11.5 & 48.6 & 19.9 & \textbf{36.3} & 34.8 \\

\midrule

\textbf{DDR-topk} & coCondenser (2.3M×8) & 61.9 & 31.0 & 13.6 & 52.8 & \underline{20.3} & \underline{34.5} & \underline{35.7} \\
\textbf{DDR-prior}  & coCondenser (2.3M×8) & \underline{65.3}  & \underline{31.4}  & \textbf{15.4} & 54.0  &    \textbf{22.3}   & \textbf{36.3}  &  \textbf{37.5}\\

\bottomrule
\end{tabular}

\label{table: primary experiment results}
\end{table*} 
\begin{table*}[h]
\centering
\caption{
 Ablation study of DDR. We remove the trainable routing and assign uniform weights to all modules (w/o routing). We then remove the prior domain labels and let DDR activate all eight modules (w/o prior).
}
\begin{tabular}{llccccccc}
\toprule 
\textbf{Method} &  \textbf{PLM} (\# Params) &  \textbf{TREC} & \textbf{NFC} & \textbf{SCD} & \textbf{SCF}  & \textbf{CLI} & \textbf{DBP} & \textbf{Avg.}  \\ 
 &   &  \colorbox{cblue}{SCI}\colorbox{cgreen}{QA} & \colorbox{cblue}{SCI}\colorbox{cgreen}{QA}  & \colorbox{cblue}{SCI}\colorbox{corange}{SUM}  & \colorbox{cblue}{SCI}\colorbox{cyellow}{FC}  & \colorbox{cyellow}{FC} & \colorbox{cred}{WIKI} & \\
\midrule

\textbf{DDR-prior}  & coCondenser (2.3M×8) & 65.3  & 31.4 & 15.4 & 54.0  &  22.3   & 36.3  & 37.5 \\
\textbf{w/o  routing} & coCondenser (2.3M×8) & 64.3 & 30.9 & 13.5 & 52.8 & 22.3 & 36.3 &  36.3\\
\textbf{w/o  prior} & coCondenser (2.3M×8)& 60.9 & 31.1 & 13.2 & 52.3 & 20.0 & 34.6 & 35.4 \\

\bottomrule
\end{tabular}

\label{table:ablation_study}
\end{table*} 

\section{Experimental results}
\label{sec:experimental_results}

To test the advantages of the DDR approach, we empirically evaluate our approach with other strong baselines in zero-shot retrieval tasks, which demand the integration of various domains to achieve effective retrieval.



\subsection{Datasets and Metrics}
\label{section: zero-shot settings}

During training, we use BERRI (Bank of Explicit Retrieval Instructions) \cite{asai2022task}, a large-scale retrieval dataset with expert-written task instructions, where each instruction is decomposed into multiple domain labels. As a result, we can extract several domains per task.
During evaluation, we test DDR on the zero-shot retrieval benchmark BEIR \cite{thakur2021beir}. Following prior work \cite{asai2022task,dai2022promptagator}, we exclude Natural Questions \cite{wagner2001natural}, MS MARCO \cite{nguyen2016ms}, HotpotQA \cite{yang2018hotpotqa}, FEVER \cite{thorne2018fever}, and CQADupStack \cite{hoogeveen2015cqadupstack} to avoid overlap between training and test tasks. We report the official NDCG@10 metric.

\subsection{Experimental Setup}
We divide training into two phases: general prefix training and domain prefix training. In the first phase, we use general retrieval tasks (e.g., MS-MARCO~\cite{nguyen2016ms}) to train only the general prefix. This helps the model learn common knowledge and ensures basic retrieval ability. In the second phase, we jointly train the domain prefix vector and the routing function. We follow the coCondenser training procedure~\cite{gao2021unsupervised} and use a pre-trained condenser-base model as the backbone. Following prior work~\cite{tang2022dptdr}, the prefix length is set to 128. We set the number of modules to 8 and set the 2 in the top-$k$ routing. The learning rate is 7e-3 for general prefix training and 7e-6 for domain prefix training. For each positive document, we select 5 negative passages as in BERRI~\cite{asai2022task}.  All training and inference are performed on a single server with 1 NVIDIA A100 80G GPU.

\subsection{Baselines}
We compare DDR with several strong retrieval baseline methods, including sparse retrieval approaches like DeepCT~\cite{dai2020context} and SPARTA~\cite{zhao2020sparta}, as well as dense retrieval approaches like Contriever~\cite{izacard2022unsupervised} and ANCE~\cite{xiong2020approximate}. We also compare DDR with a standard dense retrieval model that updates all model weights during training. We implement the DDR with various routing strategies (refer to the Sec.\ref{sec:routing}). DDR-topk uses a top-$k$ routing strategy. DDR-prior uses prior information to guide the routing strategy. 

\subsection{Experimental Results and Analysis}
\label{section: zero-shot analysis}

Table~\ref{table: primary experiment results} shows the results on the six zero-shot retrieval benchmarks in BEIR. Compared to other strong sparse retrieval and dense retrieval baselines, DDR-prior achieves the best average performance among all models, while using fewer trainable parameters. This shows its strong generalization ability and parameter efficiency. In particular, DDR-prior outperforms Contriever and ANCE by a clear margin on most datasets. Both DDR-prior and DDR-topk perform better than standard dense retrieval (DR), indicating that DDR generalizes better than DR. Compared with DDR-topk, DDR-prior improves the average score by 1.8 points, highlighting the importance of prior information for generalization.  

DDR also demonstrates interpretability compared to DR approaches. For instance, DDR shows a great improvement over DR in Fact-Checking-related tasks. This outcome suggests that the Fact-Checking retrieval module is effectively trained during the training phase. Such observations provide valuable insights for guiding the allocation of training resources to areas where they are most needed for further enhancement.

\subsection{Ablation Study}

In this subsection, we study the various components that can improve performance. First, we discard the routing function and activate the modules only based on the prior domain label. Then we assign a uniform routing to each module (w/o  routing). As shown in Table \ref{table:ablation_study}, the performance of DDR-prior will decrease by 1.2 when discarding the routing function. This result indicates the importance of routing distribution in the training of DDR. Then we remove the prior information, which activates all eight modules during retrieval (w/o prior). The performance drops by 2.1 when we activate all the modules. This demonstrates that the prior labeled domain is necessary to specialize each retriever in a specific domain. 

\section{Conclusion}
\label{conclusion}
This paper proposes a new dense retrieval paradigm, dynamic dense retrieval. We argue that DDR can address the limitations of traditional dense retrieval systems by offering flexibility, scalability, and efficiency, which makes it particularly well-suited for dynamic environments, low-resource scenarios, and applications requiring frequent updates. Specifically,  DDR has the potential to solve several long-standing challenges in dense retrieval, such as adaptability to new tasks and dynamic knowledge management. Our experiments show the advantages of this framework. It improves flexibility and efficiency while using only a small number of trainable parameters. Currently, our dense retrieval model is built on coCondenser. In future work, we plan to explore stronger backbone models trained based on large language models, such as RepLLaMA \cite{ma2024fine}.

\bibliographystyle{ACM-Reference-Format}
\bibliography{sample-base}











\end{document}